\pdfoutput=1
\documentclass[twocolumn,aps,superscriptaddress]{revtex4-2}

\usepackage{amssymb,amsfonts,amsmath,amsthm}
\usepackage{color,enumerate,graphicx}
\usepackage{hyperref}
\usepackage[normalem]{ulem}

\newcommand{\eps}{\varepsilon}
\newcommand{\R}{\mathbb R}

\newcommand{\m}{\mathbf{m}}
\newcommand{\mpa}{m_\parallel}

\newcommand{\mpe}{\mathbf{m}_\perp}

\begin{document}

\title{Theory of magnetic field-stabilized compact skyrmions
 in thin film ferromagnets}



\author{Anne Bernand-Mantel}

\affiliation{Universit\'e de Toulouse, Laboratoire de Physique et
  Chimie des Nano-Objets, UMR 5215 INSA, CNRS, UPS, 135 Avenue de
  Rangueil, F-31077 Toulouse Cedex 4, France}

\affiliation{Centre d’Elaboration de Matériaux et d’Etudes Structurales, CEMES-CNRS, 29 
rue Jeanne Marvig, 31055 Toulouse, France}

\email{anne.bernand-mantel@cemes.fr}

\author{Ana\"{i}s Fondet}

\affiliation{Universit\'e de Toulouse, Laboratoire de Physique et
  Chimie des Nano-Objets, UMR 5215 INSA, CNRS, UPS, 135 Avenue de
  Rangueil, F-31077 Toulouse Cedex 4, France}

\affiliation{Centre d’Elaboration de Matériaux et d’Etudes Structurales, CEMES-CNRS, 29 
rue Jeanne Marvig, 31055 Toulouse, France}

\author{Sarah Barnova}

\affiliation{Universit\'e de Toulouse, Laboratoire de Physique et
  Chimie des Nano-Objets, UMR 5215 INSA, CNRS, UPS, 135 Avenue de
  Rangueil, F-31077 Toulouse Cedex 4, France}

\author{Theresa M. Simon}

\affiliation{Institut für Analysis und Numerik, Westfälische Wilhelms-Universität Münster, Einsteinstr. 62, 48149 Münster, Germany}
  
\author{Cyrill B. Muratov}

\affiliation{Department of Mathematical Sciences, New Jersey Institute
  of Technology, Newark, New Jersey 07102, USA}
  
\affiliation{Departimento di Matematica, Universit\`a di Pisa, Largo
  B. Pontecorvo, 5, 56127 Pisa, Italy}


\date{\today}

\begin{abstract}
  We present a micromagnetic theory of compact magnetic skyrmions
  under applied magnetic field that accounts for the full dipolar
  energy and the interfacial Dzyaloshinskii-Moryia interaction (DMI)
  in the thin film regime. Asymptotic analysis is used to derive
  analytical formulas for the parametric dependence of the skyrmion
  size and rotation angle, as well as the energy barriers for collapse
  and bursting, two processes that lead to a finite skyrmion
  lifetime. We demonstrate the existence of a new regime at low DMI,
  in which the skyrmion is stabilized by a combination of non-local
  dipolar interaction and a magnetic field applied parallel to its
  core, and discuss the conditions for an experimental realization of
  such field-stabilized skyrmions.
\end{abstract}

\maketitle
\paragraph*{Introduction.}
Since the first real-space observation of skyrmions in a chiral magnet
\cite{muhlbauer09}, there have been a great number of theoretical and
experimental studies devoted to skyrmionics
\cite{everschor-sitte18,fert17}. These developments are motivated by
the remarkable compactness, stability and tunability of skyrmions,
which makes them attractive candidates for future applications in
information technology \cite{fert17,zhang20}.  Starting with the first
experiments in chiral magnets \cite{muhlbauer09}, it was observed that
skyrmions spontaneously appear in a certain range of non-zero magnetic
field in various bulk materials and thin-film heterostructures
\cite{tokura21}. Indeed, while the ground state of chiral magnets is
the helical state at zero magnetic field, the ferromagnetic ground
state is restored when a sufficiently strong magnetic field is
applied. However, isolated skyrmions and skyrmion lattices may exist
as metastable states under applied magnetic fields, as predicted in
the late 80's \cite{bogdanov89,bogdanov89a}. The precise dependence of
an isolated skyrmion size on the applied magnetic field was
investigated numerically
\cite{bogdanov89a,bogdanov94,bogdanov94a,bogdanov99}. In the isolated
skyrmion regime, the size of a skyrmion decreases (increases) with
increasing (decreasing) magnetic field applied antiparallel to magnetization in the
skyrmion core \cite{bogdanov89a,bogdanov94a}.

While the antiparallel configuration has been widely investigated
theoretically and experimentally
\cite{melcher14,dupe14,siemens16,buttner18,wang18,desplat20,romming13,nagaosa13,romming15,mougel20,herve18,tokura21},
the parallel configuration was considered in only a handful of studies
\cite{bogdanov94a,bogdanov99,kiselev11,bernand-mantel18}.  The reason
is that most studies on skyrmions are carried out in the strong DMI
regime, where isolated skyrmions are very easily destabilized by a
magnetic field applied parallel to the skyrmion core via the so-called
``bursting'' phenomenon, whereby the magnetic layer is reversed to the
uniform state as the skyrmion radius goes to infinity.  The position
of the bursting line on the skyrmion stability diagram was estimated
by numerical minimization in previous theoretical works
\cite{bogdanov94a,bogdanov99,kiselev11,bernand-mantel18}.  A solution
to avoid this instability is to confine the skyrmion in a dot to
extend its stability to a wider range of applied fields
\cite{tejo18,winkler21,cortes-ortuno19}. Another possibility is to
work in the low DMI regime. In this regime, it was recently
demonstrated \cite{bernand-mantel20} that the non-local dipolar
interaction that has been classically neglected
\cite{bogdanov94a,bogdanov99,lobanov16}, may become comparable to the
DMI and play a role in skyrmion stabilization. Several works based on
approximate models suggest that the non-local dipolar interaction
modifies the range of existence of isolated skyrmions in applied
magnetic fields \cite{kiselev11,bernand-mantel18}, but this regime has
remained largely unexplored.

In this Letter, we carry out an investigation of isolated compact
skyrmions under an applied magnetic field in the thin film and low DMI
regime, taking into account the full stray field. We carry out an
asymptotic analysis and obtain analytical formulas predicting the
skyrmion radius, angle, as well as the collapse and bursting energy
barriers as functions of the system parameters. We obtain a skyrmion
stability diagram, which reveals a different response to a parallel
field at low DMI strength compared to the existing phase diagrams
\cite{bogdanov99,kiselev11,bernand-mantel18}. This modification of the
skyrmion phase diagram due to stray field effects is confirmed by
direct micromagnetic simulations. We emphasize that in this regime the
parallel field enables to increase the skyrmion size and stability and
discuss possible routes to observe field-stabilized skyrmions
experimentally in conventional skyrmion materials.


\paragraph*{Model.} We consider a ferromagnetic thin film with
perpendicular magnetic anisotropy and interfacial DMI. Following our
previous works \cite{ms:prsla16,kmn:arma19,m:cvar19,bernand-mantel20}
with the addition of the Zeeman energy term, the micromagnetic energy
of a magnetization $\m : \R^2 \to \mathbb S^2$ reduces to (see
\cite{suppl} for details):
\begin{align}
  \label{E}
  \begin{split}
    E(\m) & = \int_{\R^2} \left( |\nabla \m|^2 + (Q - 1) |\mpe|^2
    \right) d^2 r \\ & - \int_{\R^2} \left( 2
  h(m_\| + 1) + 2 \kappa
  \mpe \cdot \nabla \mpa \right) \, d^2 r  \\
    & - {\delta \over 8 \pi} \int_{\R^2} \int_{\R^2} {(\mpa(\mathbf r)
      - \mpa(\mathbf r'))^2 \over |\mathbf r - \mathbf r'|^3} \, d^2 r
    \, d^2 r' \\ & + {\delta \over 4 \pi} \int_{\R^2} \int_{\R^2}
    {\nabla \cdot \mpe(\mathbf r) \, \nabla \cdot \mpe (\mathbf r')
      \over | \mathbf r - \mathbf r'|} \, d^2 r \, d^2 r',
  \end{split}
  \end{align}
  where $\mpe \in \R^2$ is the in-plane component of $\m$, i.e., the
  component perpendicular to the out-of-plane anisotropy easy axis,
  and $\mpa \in \R$ is the out-of-plane component of $\m$, i.e., the
  component parallel to the easy axis.  The energy is measured in the
  units of $Ad$, where $A$ is the exchange stiffness and $d$ the film
  thickness. Lengths are measured in the units of the exchange length
  $\ell_\mathrm{ex} = \sqrt{2A /(\mu_0 M_\mathrm{s}^2)}$, where
  $M_\mathrm{s}$ is the saturation magnetization and $\mu_0$ the
  vacuum magnetic permeability.  The dimensionless film thickness is
  $\delta = d / \ell_\mathrm{ex} \lesssim 1$.  We assume that the
  external magnetic field is applied normally to the film plane. We
  introduced the dimensionless quality factor
  $Q = K_{\mathrm{u}} / K_{\mathrm{d}} > 1$, where $K_\mathrm{u}$ is
  the magnetocrystalline anisotropy constant and
  $K_{\mathrm{d}}= \frac12 \mu_0 M_\mathrm{s}^2$, the dimensionless
  DMI strength $\kappa = D / \sqrt{AK_{\mathrm{d}}}$, and the
  dimensionless applied magnetic field $h$ defined as
  $h= (\mathbf H \cdot \hat{\mathbf z}) / M_s$.  The first four energy
  terms are local and represent, respectively, the exchange energy,
  the effective anisotropy energy (magnetocrystalline energy
  renormalized to take into account the local stray field
  contribution), the Zeeman energy and the DMI energy. The last two
  terms correspond to the long-range part of the dipolar energy, which
  splits into surface and volume contributions (see \cite{suppl} for
  details).

  We are considering an isolated skyrmion with a magnetization
  pointing up in its center, in a background where the magnetization
  is pointing down. As a consequence, we assume that the applied
  magnetic field in the positive field direction is lower than the
  anisotropy field, $h < Q - 1$, to ensure that the uniform state
  $\m_0 = -\hat{\mathbf z}$ is a local minimizer of the energy.  Under
  the condition that $\m(\mathbf r) \to - \hat{\mathbf z}$
  sufficiently fast as $|\mathbf r| \to \infty$, we consider the
  skyrmion number
  $q(\m) = \frac{1 }{ 4 \pi} \int_{\R^2} \m \cdot \left(
    \frac{\partial \m }{ \partial x} \times \frac{\partial \m }{
      \partial y} \right) \, dx \, dy$ \cite{nagaosa13,braun12}, so
  that with our convention we have $q(\m) = 1$ for a skyrmion profile.

    \begin{figure}
  \centering
\includegraphics[width=5.5cm]{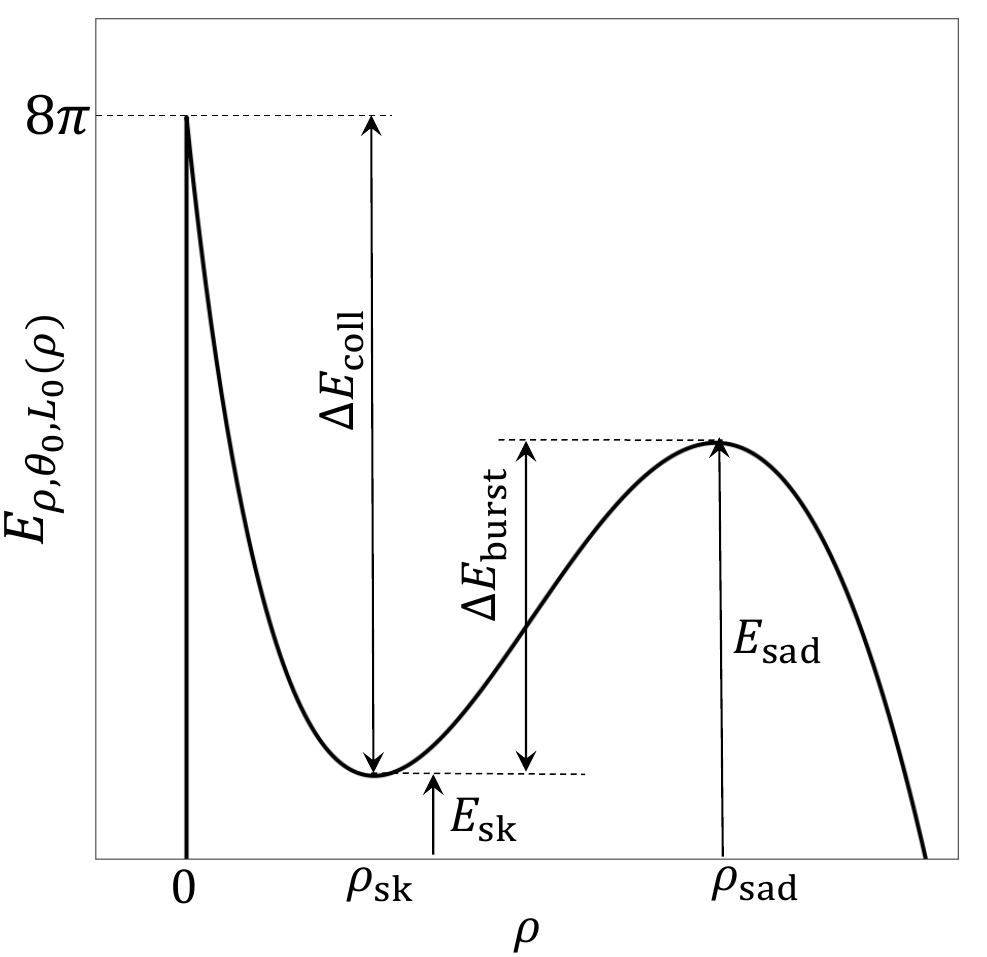}
\caption{Leading order skyrmion energy $E_{\rho, \theta, L}$ vs.
  radius $\rho$ after minimization over $\theta$ and $L$ for
  $\kappa = 0.35$, $\delta = 0.7$, $h = 0.008$ and $Q = 1.591$ (see
  (2.27) in \cite{suppl}).}
  \label{Energies}
\end{figure}

\paragraph*{Skyrmion profiles.} In the regime, in which the other
energy terms remain perturbations to the dominating exchange energy,
the skyrmion profile can be shown to be close to a Belavin-Polyakov
profile \cite{bms:arma21}, i.e., a minimizer of the exchange energy
among all $\m$ with $q(\m) = 1$ \cite{belavin75}. Therefore, we can
proceed as in \cite{bernand-mantel20,bms:arma21} with an asymptotic
analysis based on a suitably truncated Belavin-Polyakov type profile
$\m_{\rho, \theta, L}$ (for the precise form, see \cite{suppl}), with
the necessary modifications to account for the presence of the applied
field. The equilibrium radius $\rho$, the rotation angle $\theta$ and
the cutoff scale $L$ are obtained \cite{bms:arma21,suppl} from the
minimization of the leading order energy
$E(\m_{\rho, \theta, L}) \simeq E_{\rho,\theta,L}$, where (see
Fig.~\ref{Energies} for an illustration):
  \begin{multline}
    \label{eq:ErtL}
    E_{\rho,\theta,L} = 8\pi + \frac{4\pi}{L^2} + 4\pi (Q-1 -
    h)\rho^2 \log \left( \frac{4 L^2}{e^{2(1+\gamma)} }\right)
    \\
    \quad - 4 \pi h \rho^2 - 8\pi \kappa \rho \cos \theta +
    \frac{\pi^3 \rho \delta}{8} \left( 3\cos^2 \theta - 1\right),
  \end{multline}
  where $\gamma \approx 0.5772$ is the Euler-Mascheroni constant.
  Under the assumption $ \bar \eps \ll 1$, where
\begin{align}
\label{eps}
   \bar \eps = \frac{1 }{ \sqrt{1 - \bar h}} \times
		\begin{cases}
                  \left(8\pi |\bar \kappa| - \frac{\pi^3}{4} \bar \delta \right)
                  & \text{ if } |\bar\kappa|\geq
                  \frac{3\pi^2}{32}\bar\delta,\\
                  \left(\frac{128\bar\kappa^2}{3\pi\bar\delta} +
                    \frac{\pi^3}{8}\bar\delta \right) & \text{ else,}
 		 \end{cases}
\end{align} 
with
\begin{align}
\label{var2}
  \bar \delta = {\delta \over \sqrt{Q - 1}}, \qquad \bar \kappa
  = {\kappa \over 
  \sqrt{Q - 1}},  \qquad
  \bar h = {h \over Q - 1},
\end{align}
the minimizer of the reduced energy in \eqref{eq:ErtL} gives
  the leading order skyrmion equilibrium angle
\begin{align}
  \label{theta01}
  \theta_0 =
  \begin{cases}
    0 & \text{ if } \bar\kappa \geq \frac{3\pi^2}{32}\bar\delta,\\
    -\pi & \text{ if } \bar\kappa \leq
    -\frac{3\pi^2}{32} \bar\delta,\\
    \pm \arccos\left(\frac{32\bar\kappa}{3\pi^2\bar\delta} \right) & \text{
      else,}
  \end{cases}
\end{align}
and radius $\rho_\mathrm{sk} = \bar \rho_\mathrm{sk} / \sqrt{Q - 1}$,
in units of $\ell_\mathrm{ex}$, where
\begin{align}
  \label{rhoskyr}
  \bar \rho_\mathrm{sk} = \frac{1}{16\pi \sqrt{1 -\bar h}} \times
  \frac{\bar \eps}{ \left(-W_{-1}\left(-\beta \bar
  \eps
  \right) \right) },
\end{align}
provided that $\beta \bar \eps < e^{-1}$ (see Sec. 2.3 in
\cite{suppl}). Here $W_i$ refers to the $i$-th real-valued branch of
the Lambert $W$ function \cite{corless96}, and
$\beta = {e^{1 + \gamma} \over 32 \pi} \exp \left( {\bar h \over 2 (1
    - \bar h) } \right)$. The energy of the optimal skyrmion solution
is
\begin{align}
 \label{Eskyrm}
  E_\mathrm{sk} = 8\pi - \frac{ {\bar\eps}^2
  }{32\pi W_{-1}^2\left(-\beta \, \bar\eps
  \right)} \left(-W_{-1}\left(-\beta\, \bar\eps 
  \right) - \frac{1}{2}\right).
\end{align}

For $0 < 1 - \bar h \ll 1$ with $\beta \bar \eps < e^{-1}$ fixed,
there is also a minimum energy saddle point solution with equilibrium
angle $\theta_0$ from \eqref{theta01} and radius
$\rho_\mathrm{sad} = \bar \rho_\mathrm{sad} / \sqrt{Q - 1}$, where
\begin{align}
  \label{rhosaddle}
 \bar \rho_\mathrm{sad} = \frac{1}{16\pi \sqrt{1 - \bar h}} \times \frac{\bar
  \eps}{ \left(-W_{0}\left(-\beta \bar
  \eps
  \right) \right) }. 
\end{align}
The energy of the saddle is given by
\begin{align}
  \label{Esaddle}
  E_\mathrm{sad} = 8\pi - \frac{ {\bar\eps}^2}
  {32\pi W_{0}^2\left(-\beta \, \bar \eps
  \right)} \left(-W_{0}\left(-\beta\, \bar \eps
  \right) - \frac{1}{2}\right).
\end{align}

The above formulas become asymptotically exact when $\bar \eps \to 0$
and $\bar h \to 1^-$ with $\beta \bar \eps$ fixed, which corresponds
to
${\bar \delta + |\bar \kappa| \over \sqrt{1 - \bar h} } \exp \left( {1
    \over 2 (1 - \bar h)} \right) = O(1)$ as
$\bar \delta, \bar \kappa \to 0$.  Asymptotically, the skyrmion and
the saddle point solutions disappear via a saddle-node bifurcation at
a critical value of $|\bar \kappa| = \bar \kappa_c$ given by
  \begin{align}
    \label{eq:hc}
    \bar \kappa_c =
    \begin{cases}
      {\pi^2 \over 32} \bar \delta + {4 \sqrt{1 - \bar h} \over e^{2 +
          \gamma}} e^{-{\bar h \over 2 ( 1 - \bar h)}} & \text{if} \
      \bar \delta \leq \bar \delta_c, \vspace{1mm} \\
      \sqrt{{3 \pi^2 \bar \delta \over 4} \left( {\sqrt{1 - \bar h}
            \over e^{2 + \gamma}} e^{-{\bar h \over 2 ( 1 - \bar h)}}
          - {\pi^2 \bar \delta \over 256} \right)} & \text{else},
    \end{cases}
  \end{align}
  where
  $\bar \delta_c = {64 \sqrt{1 - \bar h} \over \pi^2 e^{2 + \gamma}}
  e^{-{\bar h \over 2 ( 1 - \bar h)}}$, as the value of
  $|\bar \kappa|$ is increased.

\begin{figure}[!ht]
  \centering
  \includegraphics[width=8.5cm]{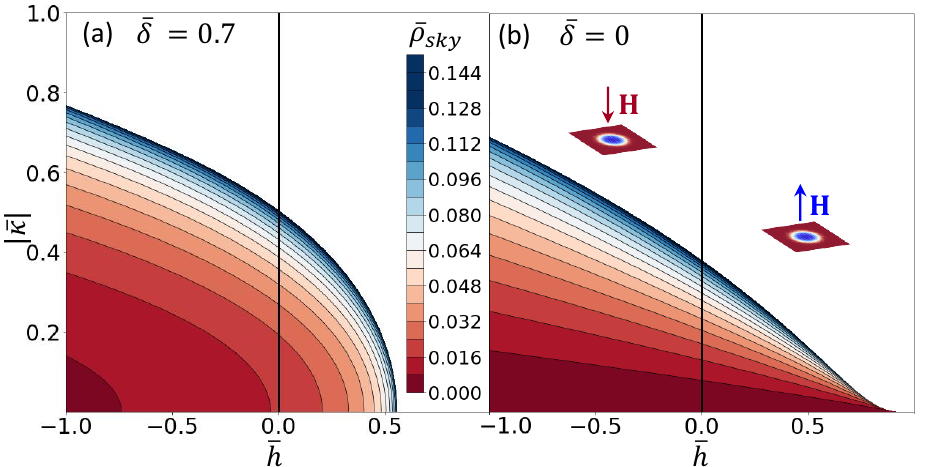}
  \caption{Scaled skyrmion radius $\bar \rho_\mathrm{sk}$ for the
    intermediate thickness regime with $\bar\delta=0.7$ (a) and
    ultrathin regime with $\bar\delta=0$ (b) as a function of
    $\bar\kappa$ and $\bar h$.  The convention for the field direction
    is shown schematically in the insets, where in the skyrmion image,
    the red and blue colors represent, respectively, down and up
    magnetizations and white corresponds to in-plane magnetization.
    For $\bar h < 0$, the field is applied antiparallel to the
    skyrmion core (red arrow in the inset), while for $\bar h >0$ the
    field is parallel (blue arrow in the inset).}
  \label{comparaison}
\end{figure}

  \paragraph*{Skyrmion phase diagram.} In Fig.~\ref{comparaison}, we
  present the asymptotic skyrmion phase diagram by plotting the scaled
  skyrmion radius $\bar \rho_\mathrm{sk}$ from \eqref{rhoskyr} as a
  function of the dimensionless DMI strength $\bar\kappa$ and applied
  field $\bar h$, where the positive field direction is parallel to
  the skyrmion core [see the insets in
  Fig.~\ref{comparaison}(b)]. Skyrmion solutions are predicted to
  exist when $0 < |\bar \kappa| + \bar \delta \lesssim 1$ and
  $\bar h \leq \bar h_c$, where $\bar h_c$ is obtained by solving for
  $h$ in \eqref{eq:hc}, while no skyrmion solutions are expected for
  $\bar h > \bar h_c$ (see also \cite{suppl}).

According to Fig.~\ref{comparaison}, in the antiparallel
configuration, $\bar h < 0$, the skyrmion radius is strongly
increasing with $|\bar \kappa|$ and increasing slower with a decrease
in $|\bar h|$. In the parallel configuration, $\bar h > 0$, in the
intermediate thickness ($\bar \delta = 0.7$) regime shown in
Fig.~\ref{comparaison}(a), the skyrmion radius dependence is, to the
contrary, dominated by its magnetic field dependence and the
dependence on $|\bar \kappa|$ is strongly reduced for small
$|\bar \kappa|$. In particular, the lines of equal radius present a
concave character and become parallel to the $y$ axis as
$|\bar \kappa| \to 0$. This diminished dependence of
$\bar \rho_\mathrm{sk}$ on $|\bar \kappa|$ comes from the fact that
the dominant stabilization mechanism in the low $|\bar \kappa|$ regime
is the long-range dipolar interaction. The skyrmion behavior in this
regime presents remarkable differences from what is predicted by the
classical skyrmion theory, which only takes into account the local
dipolar interaction \cite{bogdanov94a,bogdanov99,lobanov16}. For
comparison, we set $\bar \delta =0$ in our model in
Fig.~\ref{comparaison}(b), which is equivalent to neglecting the
long-range dipolar interactions.  In this case, the lines of constant
radius do not present this strong concave character and the main
source of skyrmion stabilization in the whole field range is solely
the increase of $|\bar \kappa|$.
\begin{figure}[h!]
  \centering
  \includegraphics[width=8.5cm]{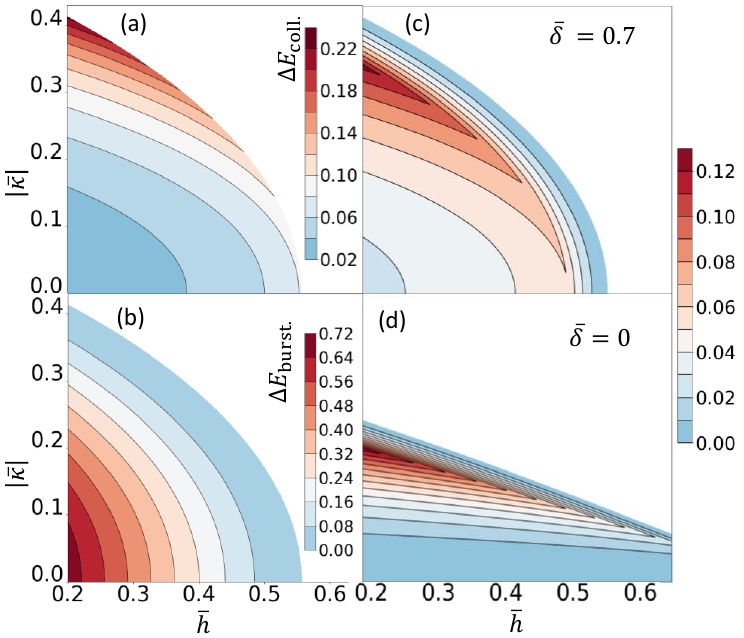}
  \caption{Skyrmion collapse $\Delta E_\mathrm{coll}$ (a) and bursting
    $\Delta E_\mathrm{burst}$ (b) energy barriers (in units of $Ad$)
    for the intermediate thickness regime with $\bar\delta=0.7$.  Also
    shown is the effective barrier
    $\Delta E = \min (\Delta E_\mathrm{coll}, \Delta
    E_\mathrm{burst})$, which illustrates the optimal stability region
    against collapse and bursting for $\bar\delta=0.7$ (c) and for the
    ultrathin film with $\bar\delta=0$ (d).}
  \label{collandburst}
\end{figure}
\paragraph*{Skyrmion stability.} A second important outcome of our
analysis is the prediction of the parametric dependence of the
bursting energy barrier.  Previous studies on skyrmion stability have
focused on the collapse phenomenon and the field applied antiparallel
to the skyrmion core \cite{melcher14,bessarab15,buttner18,
  bernand-mantel20,cortes-ortuno17,desplat20,hoffmann20}. However, for
a field applied parallel to the skyrmion core, bursting is the main
source of instability.  These two energy barriers are shown in
Fig.~\ref{Energies}, where we present the skyrmion energy as a
function of the  radius $ \rho $ for a
given set of parameters (see (2.27) in \cite{suppl}).  The local
minimum of energy at $\rho_\mathrm{sk}$ corresponds to the
equilibrium skyrmion solution. For $ \rho<  \rho_\mathrm{sk}$
the energy is increasing with decreasing $\rho$ and reaches the
value $8\pi$ at $\rho = 0$, as all the energies tend to zero except
the exchange energy, which tends to its minimum value of $8 \pi$ among
configurations with $q = 1$, as first demonstrated in \cite{belavin75}
and discussed previously
\cite{melcher14,buttner18,bernand-mantel18,bernand-mantel20,bms:arma21}.

The energy difference between the equilibrium skyrmion and the ``zero
radius skyrmion'' is thus
$\Delta E_\mathrm{coll} = 8 \pi - E_\mathrm{sk}$, which serves as the
energy barrier to skyrmion collapse \cite{bernand-mantel22}.  As the
radius becomes larger than $\rho_\mathrm{sk}$, the energy is
increasing and a saddle point of the micromagnetic energy is observed
at $\rho_\mathrm{sad}$. This saddle prevents the skyrmion solution
from bursting as the skyrmion energy would otherwise go to $-\infty$
as the skyrmion radius tends to infinity. The bursting barrier is
defined as $\Delta E_\mathrm{burst} = E_\mathrm{sad} - E_\mathrm{sk}$.
The collapse and bursting energy barriers are plotted as functions of
$|\bar\kappa|$ and $\bar h$ in Figs.~\ref{collandburst}(a) and (b) for
the magnetic field applied parallel to the skyrmion core.
\begin{figure}[h!]
\includegraphics[width=7.5cm]{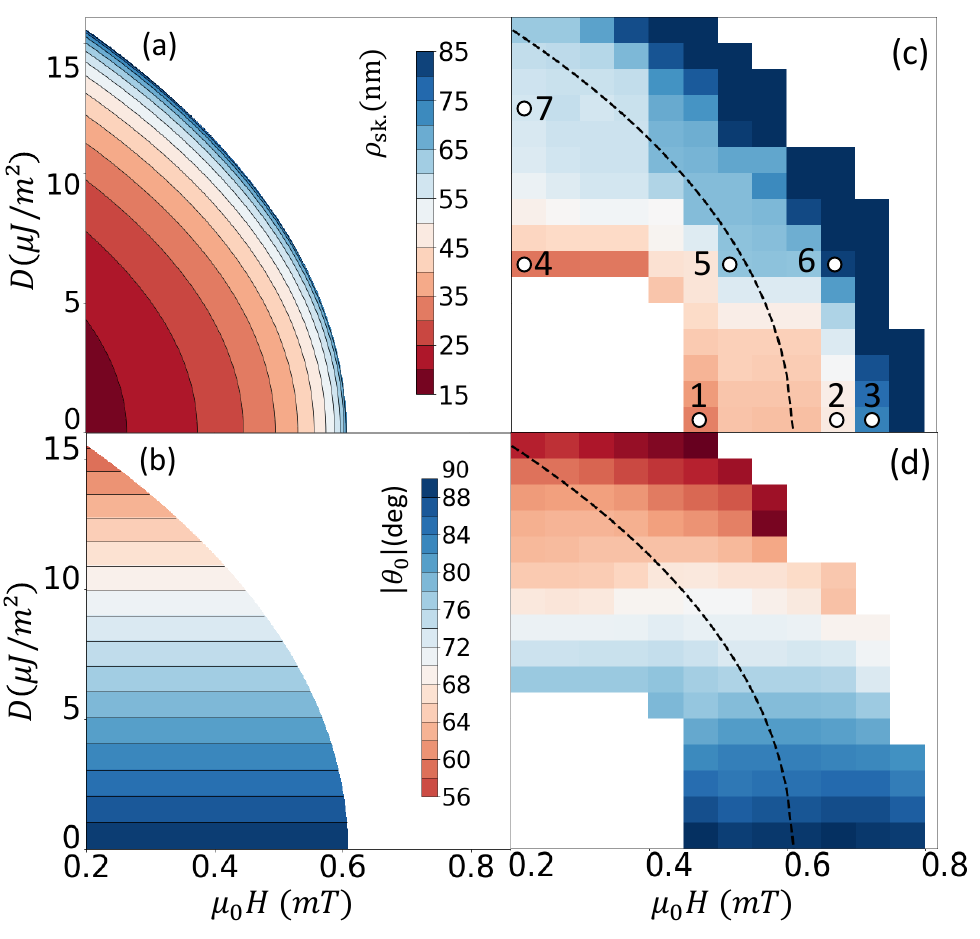}
\caption{The skyrmion characteristics in the low DMI regime for
    $d = 5$ nm, $A=20$ pJ/m, $M_{\mathrm{s}}= 10^5$ A/m, and
    $K_\mathrm{u} = 6346$ J/m$^3$ corresponding to $Q = 1.01$: (a)
  Skyrmion radius $r_{\mathrm{sky}}$ and (b) skyrmion rotation
  angle $|\theta_0|$ from \eqref{rhoskyr} and \eqref{theta01}; (c)
    skyrmion radius and (d) skyrmion rotation angle from MuMax3
    simulations on a $2048 \times 2048$ nm$^2$ square box with mesh
  size $2 \times 2 \times 5$ nm. The dashed line is the line of zero
  bursting barrier defined in \eqref{eq:hc}. }
\label{mumax}
\end{figure}
The collapse and bursting energy barriers present opposite variations
as the effective DMI or applied magnetic field are increased: an
increase of $|\bar\kappa|$ or $\bar h $ increases the collapse
barrier, but decreases the bursting barrier. As a consequence, in the
positive field regime, our theory predicts that the optimum stability
region for skyrmions is where both the collapse and the bursting
barriers are large. This optimum stability region is illustrated in
Figs.~\ref{collandburst}(c) and (d), where we plot the effective
energy barrier
$\Delta E = \min (\Delta E_\mathrm{coll}, \Delta E_\mathrm{burst})$.
In Fig.~\ref{collandburst}(c), for an intermediate thickness
($\bar \delta =0.7$), the optimal stability is obtained at large
$ |\bar\kappa|$, for low applied fields.  As $ |\bar\kappa|$ is
decreased, the region of optimum stability, which appears in reddish
colors, is shifted to positive fields, showing that the applied
positive field can partly compensate the decrease of DMI to stabilize
skyrmions in the low $|\bar\kappa|$ regime. The stabilization of the
skyrmion in this regime is due to a combined effect of the applied
positive field and the long-range dipolar interaction.  In
Fig.~\ref{collandburst}(d), we present the ultrathin film limit
($\bar \delta =0$).  In that case, the positive field cannot
compensate the decrease of $|\bar\kappa|$ and hardly any increase of
skyrmion stability with applied positive field is observed in the low
$|\bar\kappa|$ regime.  Comparison between these two cases in
Figs.~\ref{collandburst}(c) and (d) reveals that the long-range
dipolar interaction provides a stabilization mechanism in low DMI
systems leading to a new regime of parallel field stabilization of
skyrmions.

\begin{figure}[h!]
\includegraphics[width=7.5cm]{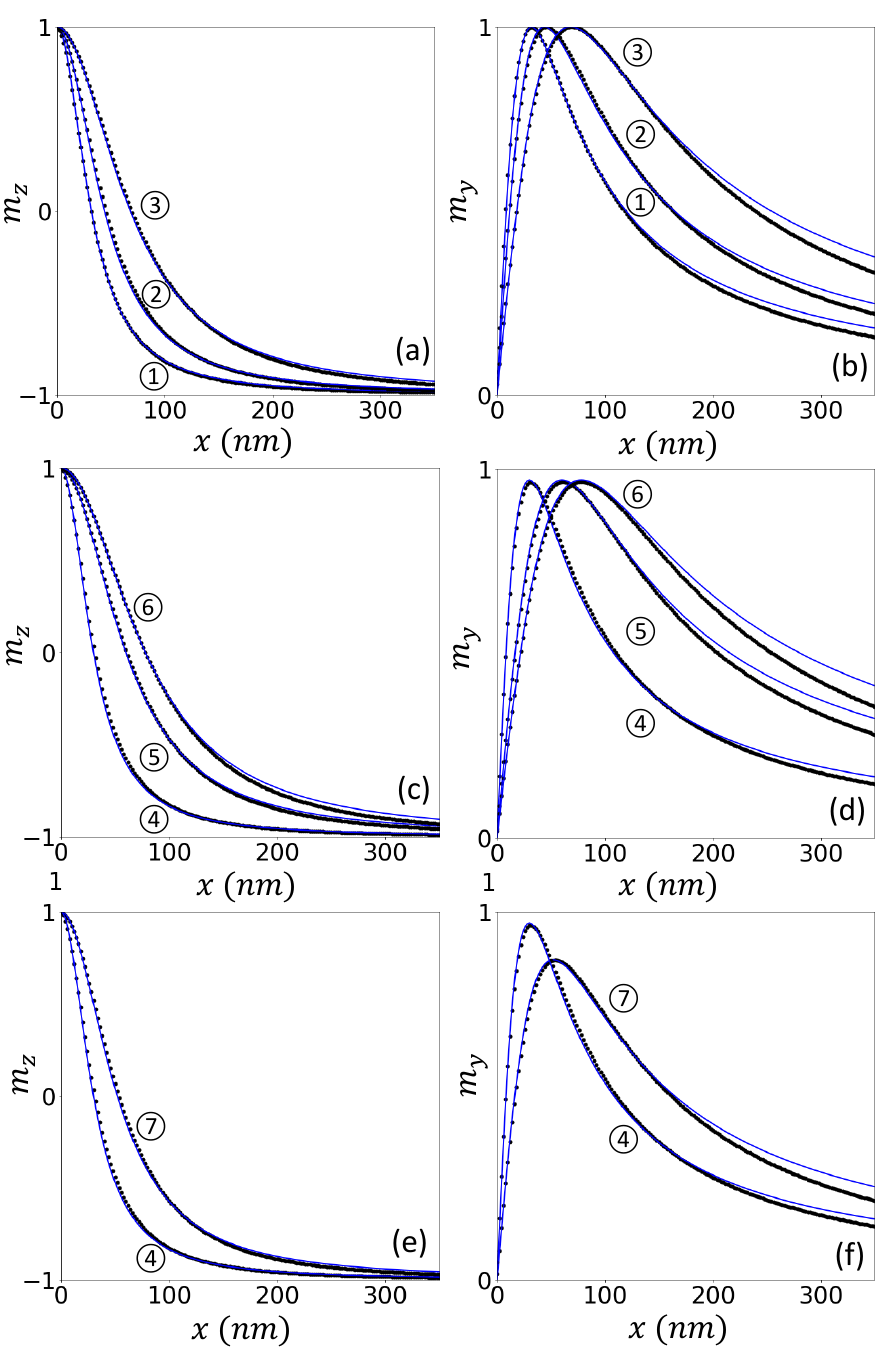}
\caption{The magnetization profiles along the $x$-axis with the
    skyrmion at the origin obtained from the MuMax3 simulations for
    the parameters indicated by the respective dots in
    Fig.~\ref{mumax}(c). The simulation datapoints are shown with the
    solid black dots, while the Belavin-Polyakov profiles obtained by
    fitting their radius and angle to the simulation data are shown
    with the solid blue lines. }
\label{BP}
\end{figure}
 \paragraph*{Micromagnetic simulations.}
 To confirm the existence of field-stabilized compact skyrmions in the
 positive field regime, we carried out micromagnetic simulations,
 using MuMax3 \cite{vansteenkiste14} (see \cite{suppl} for details on
 the procedure). The chosen dimensional parameters correspond to a
 ferrimagnetic materials with a low $ M_\mathrm{s}$ and
 $ K_\mathrm{u}$ (see Fig.~\ref{mumax} caption). In Fig.~\ref{mumax},
 we present the skyrmion radius (a) and rotation angle (b) predicted
 by our model [see \eqref{rhoskyr} and \eqref{theta01}] for this set
 of parameters. The corresponding values of the skyrmion radius and
 angle from the simulations are presented, respectively, in
 Figs.~\ref{mumax}(c) and (d), where the dashed line is the line of
 zero bursting barrier defined in \eqref{eq:hc}. Figure~\ref{mumax}(c)
 also shows several points in the parameter space for which we present
 the numerical magnetization profiles in Fig.~\ref{BP}. These profiles
 agree very well with the suitably rotated and dilated
 Belavin-Polyakov profiles, consistently with our theoretical
 predictions.

 The simulations and the theory show a good agreement: they both
 predict, at fixed DMI, an increase of the skyrmion radius from around
 20 nm up to bursting as the positive field increases
 [Figs.~\ref{mumax}(a) and (c)], and at fixed magnetic field, a
 reorientation of the skyrmion angle from around $60^\circ$ to
 $90^\circ$ as $\bar\kappa $ is decreased to zero
 [Figs.~\ref{mumax}(b) and (d)]. This confirms the existence of
 skyrmions stabilized by a parallel field in the low $|\bar\kappa|$
 regime.
%

In the top right part of Figs.~\ref{mumax}(c) and (d) the white zone
corresponds to a region where the skyrmion has bursted and the system
is homogeneously magnetized in the direction of the positive applied
field. The numerical solution persists beyond the dashed line
representing the region of existence of our solution. This is expected
due to the asymptotic nature of the theory. The numerical solution
persists in the form of a less compact profile before bursting
occurs. This further extends the predicted range in which skyrmions
can be stabilized with the help of a positive field.

 \paragraph*{Conclusions and perspective.}

 We have derived the magnetic field dependence of the compact skyrmion
 size and rotation angle, along with its thermal stability, in the
 presence of DMI and full dipolar interaction. In particular, we
 obtained a condition for skyrmion bursting at positive magnetic
 field, which agrees quantitatively with micromagnetic simulations in
 the regime of compact skyrmions and which may be an invaluable tool
 for predicting this type of instability. We also established that a
 balance needs to be found between collapse and bursting energy
 barriers to optimize the skyrmion stability.

 Our analysis reveals that due to the presence of long-range dipolar
 interaction an increase of the magnetic field applied parallel to the
 skyrmion core increases the size and stability of a skyrmion
 similarly to the effect of increasing the DMI strength. A reason why
 a positive applied field has never been used to stabilize skyrmions
 so far is that the samples in which skyrmions are observed very often
 exhibit spontaneous demagnetization in the form of a helicoidal state
 or a stripe state at zero applied field. This phenomenon occurs in
 the presence of a large DMI or/and due to long-range demagnetizing
 effects. In the quest for room temperature nanometer size skyrmions,
 recent works have focused on ferrimagnetic systems, following some
 predictions of more stable skyrmions in such system with a low $M_s$
 and thicknesses of the order of $5-10$ nm
 \cite{buttner18,bernand-mantel20}. In these systems, room temperature
 stable skyrmions are observed experimentally at zero applied magnetic
 field \cite{caretta18,quessab22} and they constitute the ideal
 candidates to observe field-stabilized skyrmions in the new regime
 predicted by our theory.

 Lastly, while our study focuses on the collapse and bursting
 mechanisms in the low effective DMI and thickness regime, it would be
 interesting to extend our investigation to systems exhibiting more
 complex mechanisms of thermal instability, as in the case of systems
 with strong DMI and strong magnetocrystalline anisotropy studied
 previously in the case of no applied field (see, e.g.,
 \cite{cortes-ortuno17,desplat18,heil19,varentcova20}).

 The work of C.\ B.\ M. was supported, in part, by NSF via grant
 DMS-1908709. T.\ M.\ S. was funded by the Deutsche
 Forschungsgemeinschaft (DFG, German Research Foundation) under
 Germany's Excellence Strategy EXC 2044–390685587, Mathematics
 Münster: Dynamics–Geometry–Structure.

%


\end{document}